\documentclass[prd,
 reprint,
 amsmath,amssymb,
 aps,times
]{revtex4-2}

\usepackage{amsmath}
\usepackage{booktabs} 
\usepackage{threeparttablex}
\usepackage{graphicx}% Include figure files
\usepackage{dcolumn}% Align table columns on decimal point
\usepackage{bm}% bold math
\usepackage{amssymb}
\usepackage{times}
\usepackage{threeparttable}

\def\apjl{The Astrophysical Journal Letter}
\def\apjs{The Astrophysical Journal Supplement}

\def\aap{Astronomy and Astrophysics}

\usepackage{orcidlink}
\begin{document}

\preprint{APS/123-QED}

\title{Correlation of galactic very-high-energy gamma-ray sources detected by LHAASO with high-energy neutrinos}% Force line breaks with \\
% \thanks{A footnote to the article title}%

\author{Long-Yu Zhang}
\affiliation{Department of Astronomy, School of Physics, Huazhong University of Science and Technology, Wuhan, Hubei 430074, China}

\author{Kai Wang\,\orcidlink{0000-0003-4976-4098}}
\email{kaiwang@hust.edu.cn}
\affiliation{Department of Astronomy, School of Physics, Huazhong University of Science and Technology, Wuhan, Hubei 430074, China}%

\date{\today}% It is always \today, today,
             %  but any date may be explicitly specified

\begin{abstract}
We conducted an investigation into the possible correlation between very-high-energy (VHE, $E>$ 0.1 TeV) gamma-ray sources, as identified by the Large High Altitude Air Shower Observatory (LHAASO), and high-energy neutrino events captured by the IceCube Neutrino Observatory. For this analysis, we utilized the First LHAASO catalog, which comprises 90 VHE gamma-ray sources, and the IceCube Event Catalog of Alert Tracks (IceCat-1), which includes 348 track-like neutrino events. Through Monte Carlo simulations, we assessed the correlations between the neutrino events and LHAASO sources, achieving the best p-values of 0.237, 0.312, and 0.115, respectively, for all LHAASO sources, KM2A sources, and WCDA sources. These findings imply that the spatial coincidence between LHAASO sources and neutrino alerts is relatively weak. Furthermore, under the assumption that gamma-rays are exclusively generated through hadronic interactions, we estimated the most optimistic theoretical neutrino fluxes from LHAASO sources and compared them with the IceCube observed fluxes. The results indicate that the majority of these sources are improbable counterparts to the detected neutrinos. In conclusion, LHAASO sources could be potential galactic neutrino emitters but currently do not exhibit compelling evidence, thereby underscoring the necessity for elevated sensitivity levels in forthcoming detection instruments like IceCube-Gen2.
\end{abstract}

\maketitle

%\tableofcontents

\section{Introduction}
\label{sec:intro}

Since the operation of the Large High Altitude Air Shower Observatory (LHAASO) \citep{cao2024first}—a large extensive air shower (EAS, Extensive Air Shower) array designed for high-energy gamma-ray detection, numerous very-high-energy (VHE, $E>$ 0.1 TeV) gamma-ray sources have been observed, with the highest energy reaching 1.4 PeV \citep{cao2021discoverya,cao2021discoveryb,cao2021ultrahigh}. These VHE gamma-ray sources emitting $\sim 100\,\rm TeV$ radiation are widely considered potential PeVatrons—astrophysical objects capable of accelerating particles to PeV energy scales~\citep{tibet2021potential,albert2020hawc,abdalla2021evidence,cao2025enigmatic}. However, a critical unresolved question persists: do these VHE gamma-rays originate from high-energy protons (hadronic processes) or electrons (leptonic processes)?
Physically, high-energy protons can generate both VHE gamma-rays and neutrinos through two possible hadronic interactions: hadronuclear ($pp$) collisions with baryon backgrounds and photohadronic ($p\gamma$) interactions with radiation fields. In contrast, leptonic processes, such as inverse Compton scattering, through which high-energy electrons upscatter low-energy photons, produce VHE gamma-rays, but without accompanying neutrinos \citep{kelner2006energy}. This distinction makes neutrinos an "exclusive probe" for identifying hadronic activity in VHE gamma-ray sources. 
%For instance, Milky Way objects like supernova remnants (SNRs) and pulsar wind nebulae (PWNe)—common PeVatron candidates—possess shock environments suitable for proton acceleration and dense target materials for $pp$ interactions, theoretically enabling concurrent VHE $\gamma$-ray and neutrino emission \citep{huang2022constraints,huang2022neutrino}. Resolving the origin of LHAASO-detected VHE $\gamma$-rays thus requires linking these sources to high-energy neutrino observations.

High-energy astrophysical neutrinos have been detected for over a decade by the IceCube Neutrino Observatory~\citep{icecube2013evidence,aartsen2014observation,aartsen2015combined,aartsen2016observation, aartsen2017all, abbasi2023icecat,zegarelli2025icecat}. Although a confidence level of $\sim 4.5\sigma$ has confirmed the existence of galactic diffuse neutrinos \citep{icecube2023observation}, to date, no credible galactic neutrino individual source has been identified. 
%The nearby Seyfert galaxy NGC 1068 remains the only extragalactic object linked to neutrinos with $4.2\sigma$ confidence \citep{icecube2022evidence}. 
With the operation of many VHE gamma-ray telescopes, such as H.E.S.S. (High Energy Stereoscopic System)~\citep{aharonian2006observations}, MAGIC (Major Atmospheric Gamma Imaging Cherenkov Telescopes) \citep{aleksic2016major}, HAWC (High Altitude Water Cherenkov Observatory)~\citep{2017ApJ...843...40A}, and LHAASO, many bright VHE sources within the Milky Way galaxy have been discovered, especially 90 VHE gamma-ray sources found by LHAASO~\citep{cao2024first}. Several studies have investigated VHE gamma-ray–neutrino correlations and constrained the hadronic contribution to a subset of LHAASO sources via neutrino data~\citep{huang2022constraints,huang2022neutrino,abbasi2022searches,2024ApJ...961..254S,Alfaro_2024,li2024search}. Refs.~\citep{huang2022constraints,huang2022neutrino} searched the neutrino signals and constrained the hadronic origin of LHAASO-observed gamma rays for the first released 12 LHAASO sources that are identified with emissions above
100 TeV in 2021~\citep{cao2021ultrahigh}. Ref.~\citep{abbasi2022searches} conducted stacking
searches testing for excess neutrino emission from these 12 LHAASO sources as well as the subclasses of supernova remnants and pulsar wind nebulae, but no significant neutrino emissions were found. Ref.~\citep{2024ApJ...961..254S} evaluated the most optimistic neutrino flux for LHAASO 1908+0621, LHAASO J2018+3651, and LHAASO J2032+4102, and provided the detection expectation by IceCube and IceCube-Gen2. Ref.~\citep{Alfaro_2024} searched for a correlation between 22 known gamma-ray sources from the third HAWC
gamma-ray catalog and 14 yr of IceCube track-like data, and no significant neutrino emission from these sources was found. Ref.~\citep{li2024search} conducted searches for correlated neutrino signals from the Cygnus Bubble. In addition to the galactic sources, for extragalactic sources, similar searches have been implemented as well, e.g., radio-bright blazars~\citep{plavin2020observational,hovatta2021association}, tidal disruption events~\citep{lu2025investigating}, etc.

%Notably, Refs. \citep{2024ApJ...961..254S} and \citep{li2024search} performed spatial cross-matching and computed optimistic upper limits on hadronic neutrino fluxes—based on the full multi-year IceCube track sample, while our work focuses on the real-time IceCat-1 alert catalog.

Searching for the link between LHAASO’s VHE gamma-rays and IceCube’s neutrinos is key to identifying galactic PeVatrons and advancing our understanding of cosmic ray acceleration in the Milky Way. Along with more and more VHE galactic gamma-ray sources and neutrino events being confirmed, a systematic analysis of the correlation of neutrino events with galactic VHE gamma-ray sources becomes feasible. Unlike previous studies, we focus on a more extensive VHE source sample, i.e., the first LHAASO catalog, which includes 90 VHE $\gamma$-ray sources \citep{cao2024first}. Combining it with the released IceCube Event Catalog of Alert Tracks (IceCat-1), including 348 track-like neutrino events \citep{abbasi2023icecat}, it becomes feasible to search for the spatial coincidence between the LHAASO sources and the high-energy neutrino events ($>100\,\rm TeV$) in the skymap. A neutrino sample with a higher energy threshold, i.e., IceCat-1, is adopted since the potential PeVatrons are the most concerning research objective here. In this work, we investigate the correlation between the VHE $\gamma$-ray sources of LHAASO and the track-like neutrino events of IceCat-1, with two primary goals: (1) to determine whether the spatial overlaps between these sources and neutrinos are accidental or physically meaningful, and (2) to verify whether the LHAASO sources can explain IceCube’s observed neutrino flux under the assumption of a hadronic process. To address these goals, we first select VHE gamma-ray source samples and the neutrino samples in Section \ref{sec:sample}. We then use Monte Carlo simulations to calculate spatial coincidence p-values in Section \ref{sec:method} and compare theoretically expected neutrino fluxes by assuming that the gamma-rays detected by LHAASO are purely hadronic to the observed fluxes of IceCube in Section \ref{sec:analysis}.

\section{SAMPLE} \label{sec:sample}

\subsection{Gamma-ray source samples}

%In recent years, significant advancements in high-energy gamma-ray detection technology have driven substantial progress in relevant research. Since 2004, relying on second-generation Imaging Atmospheric Cherenkov Telescopes (IACTs) such as H.E.S.S. \citep{aharonian2006observations}, MAGIC \citep{aleksic2016major}, and VERITAS \citep{meagher2015six}, this type of detector has detected more than 200 gamma-ray sources in total.

Compared with traditional Imaging Atmospheric Cherenkov Telescopes (IACT), LHAASO has core advantages of a larger field of view and higher observation efficiency. It can achieve comprehensive observation of the VHE universe through roughly unbiased sky measurements. Meanwhile, based on the technical characteristics of the EAS array, LHAASO extends its observation energy range to the ultra-high-energy (UHE, $E \geq$ 100TeV) region, forming an overlapping coverage with the neutrino energy range detected by the IceCube Neutrino Observatory.

The detection tasks of LHAASO are undertaken by two components: the Kilometer Squared Array (KM2A) and the Water Cherenkov Detector Array (WCDA), each covering different gamma-ray energy intervals. KM2A is responsible for detecting gamma rays above 10 TeV, while WCDA focuses on TeV-level gamma rays. The combination of the two realizes complete coverage of gamma rays in the energy range from $\sim 200$ GeV to more than PeV.

Since the release of the first LHAASO catalog \citep{cao2024first}, 90 gamma-ray sources—each with an extension scale of less than $2^{\circ}$—have been identified in its all-sky map. These sources exhibit distinct distribution characteristics: 65 of them show an extended morphology, supported by a detection confidence level greater than $3\sigma$; 54 can be detected simultaneously by both the WCDA and KM2A, which confirms their observational reliability; 43 achieve a detection confidence level exceeding $4\sigma$ for energies above 100 TeV, qualifying them as typical ultra-high-energy radiation sources; and 82 fall within a galactic latitude range of $|\rm b|~<~12^{\circ}$, meaning they are concentrated in the vicinity of the galactic plane.

\subsection{Neutrino sample}

The IceCube Neutrino Observatory, situated at the South Pole, is a large-scale detector designed to detect astrophysical neutrinos using Cherenkov light \citep{abbasi2009icecube}. The optical emissions it captures are categorized into two types: "track events" and "cascade events". Since 2016, IceCube has been sharing real-time notifications of high-energy neutrino events (with energies exceeding 100 TeV) with the multi-messenger observation community via platforms such as the Global Cycling Network (GCN) \footnote{https://gcn.nasa.gov/} \citep{aartsen2017all}. In 2019, a significant upgrade was implemented to this real-time notification system \citep{blaufuss2019next}, which not only improved signal purity, improved background rejection capabilities, and expanded the range of alert coverage, but also introduced a two-level classification system ("Gold" and "Bronze") based on signal purity. These two categories correspond to astrophysical origin probabilities of $50\%$ and $30\%$, respectively, assuming that an astrophysical neutrino energy spectrum follows a power-law with an index of 2.19.

For track-like events, the angular resolution can reach better than 0.5$^\circ$ at energies above 10 TeV, whereas for cascade-like events, the angular resolution is typically on the order of 5$^\circ$ under the same reconstruction framework \citep{bukhari2024icecube}. This study focuses on "track-like neutrino candidates" due to their significantly better angular resolution compared to cascade events. IceCube has released the first catalog of track-like alert events \citep{abbasi2023icecat}. The updated (IceCube Event Catalog of Alert Tracks) IceCat-1 contains 348 track events that meet the alert criteria between May 2011 and October 2023, including 172 “Bronze” events and 176 ”Gold” events. To investigate the correlation between LHAASO sources and astrophysical neutrinos, it is necessary to analyze their spatial position. We notice that the spatial region of some neutrinos in the catalog overlaps with LHAASO sources. 
However, when a large number of sources and neutrino events are distributed in the sky, the probability of their overlap is clearly present. Therefore, we need to know whether this observational phenomenon is a spatial coincidence or a correlation between the sources and neutrino events. 

In some previous studies \citep{plavin2020observational,hovatta2021association,lu2025investigating}, some selection criteria are adopted to exclude a portion of neutrino events. In this work, we follow the same approach and perform strict selection on the IceCat-1 sample as in Ref.~\citep{plavin2020observational}. To explore the potential PeVatron candidates, we impose a neutrino energy cut of $E_\nu \geq 100\ \rm TeV$ (the parent proton energy $E_p\sim 20 E_\nu$). We primarily adopt $\Omega_{90}\leq 10 \rm  \deg^2$ as the main selection criterion for our baseline analysis. Additional thresholds of $15\rm \ deg^2$ and $20\rm \ deg^2$ are also applied to test the robustness of our results against different angular uncertainty cuts. This criterion is designed to filter out poorly localized neutrino events with large angular uncertainties. Adopting multiple thresholds also allows us to test the robustness of our results. All selection rules covered candidates from both Gold and Bronze channels. Finally, we obtain 202, 234, and 249 qualified alert events corresponding to the three $\Omega_{90}$ thresholds of $10\ \mathrm{deg^2}$, $15\ \mathrm{deg^2}$, and $20\ \mathrm{deg^2}$, respectively. All sample selection criteria are fixed in advance before all data analysis, with no post-hoc tuning. The spatial distribution, spectral characteristics, and directional uncertainties of these events provide a core data foundation for subsequent spatial coincidence analysis and flux comparison.

\section{LHAASO source-neutrino coincidence searching} \label{sec:method}

To explore the potential spatial correlation between LHAASO sources and neutrinos, we conducted a systematic search across the full sky map of LHAASO-detected sources. LHAASO sources are classified into point-like and extended sources, with their spatial boundaries defined by two key parameters from the LHAASO collaboration: $\sigma_{p,95}$, the $95\%$ statistical uncertainty of the source component’s position, and $r_{39}$, the $39\%$ containment radius of the 2D-Gaussian model used to fit the source’s spatial profile \citep{cao2024first}. For each track-like neutrino event in our sample, a $90\%$ positional uncertainty region is available from the IceCube catalog \citep{abbasi2023icecat}, which accounts for both statistical and systematic errors in event localization. The core of our spatial correlation search lies in identifying overlaps between the extended range of LHAASO sources and the $90\%$ uncertainty regions of neutrino events.

For consistency across source types, we define the source’s effective extension radius as $\sqrt{\sigma_{p,95}^2 + r_{39}^2}$, a combined metric that incorporates both positional uncertainty and intrinsic spatial extent. In this work, we do not distinguish point-like and extended LHAASO sources during the spatial search, and the uncertainty regions for related LHAASO sources are adopted as presented in the first LHAASO catalog~\citep{cao2024first}. A spatial coincidence is defined as any intersection between the effective extension of a LHAASO source and the $90\%$ uncertainty region of a neutrino alert event.

\begin{figure*}
\centering
\includegraphics[width=0.45\textwidth]{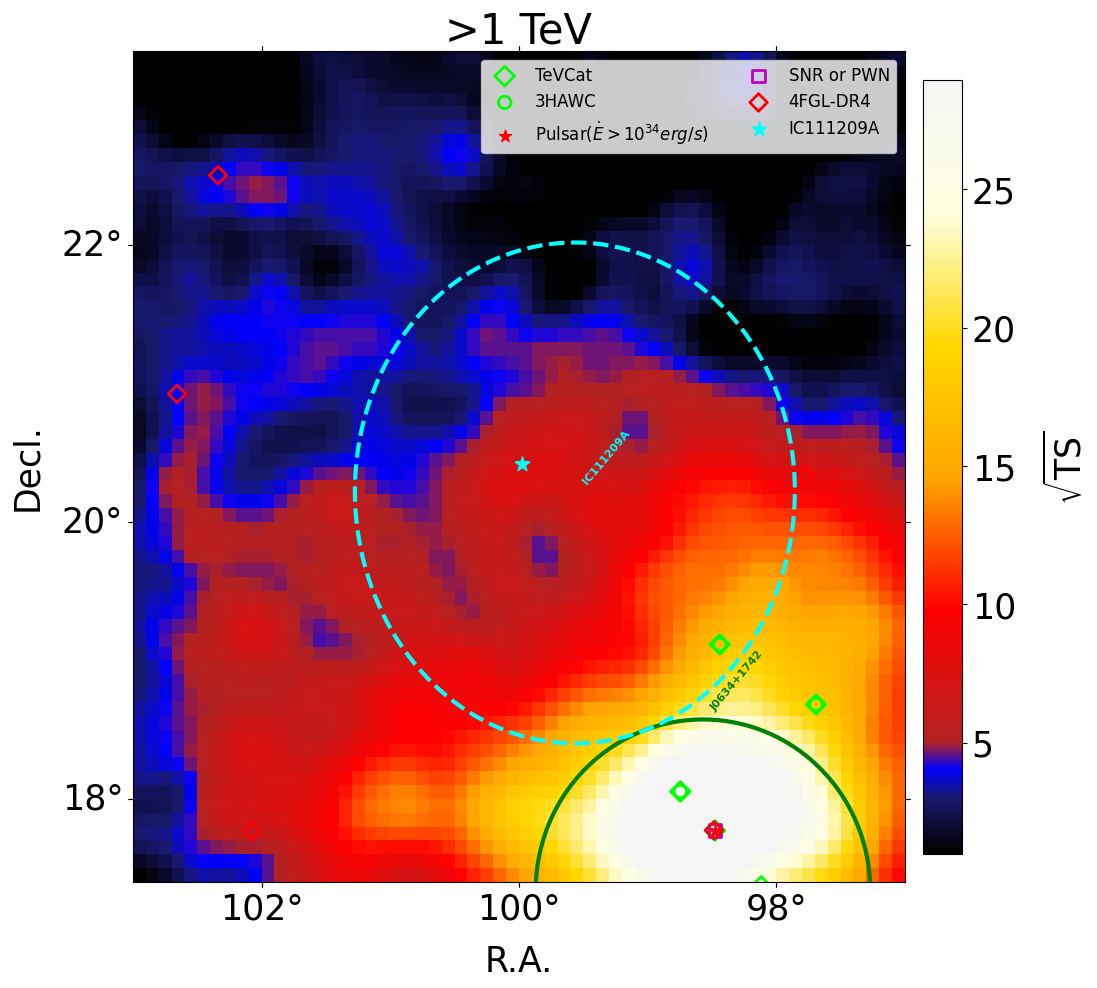}
\includegraphics[width=0.45\textwidth]{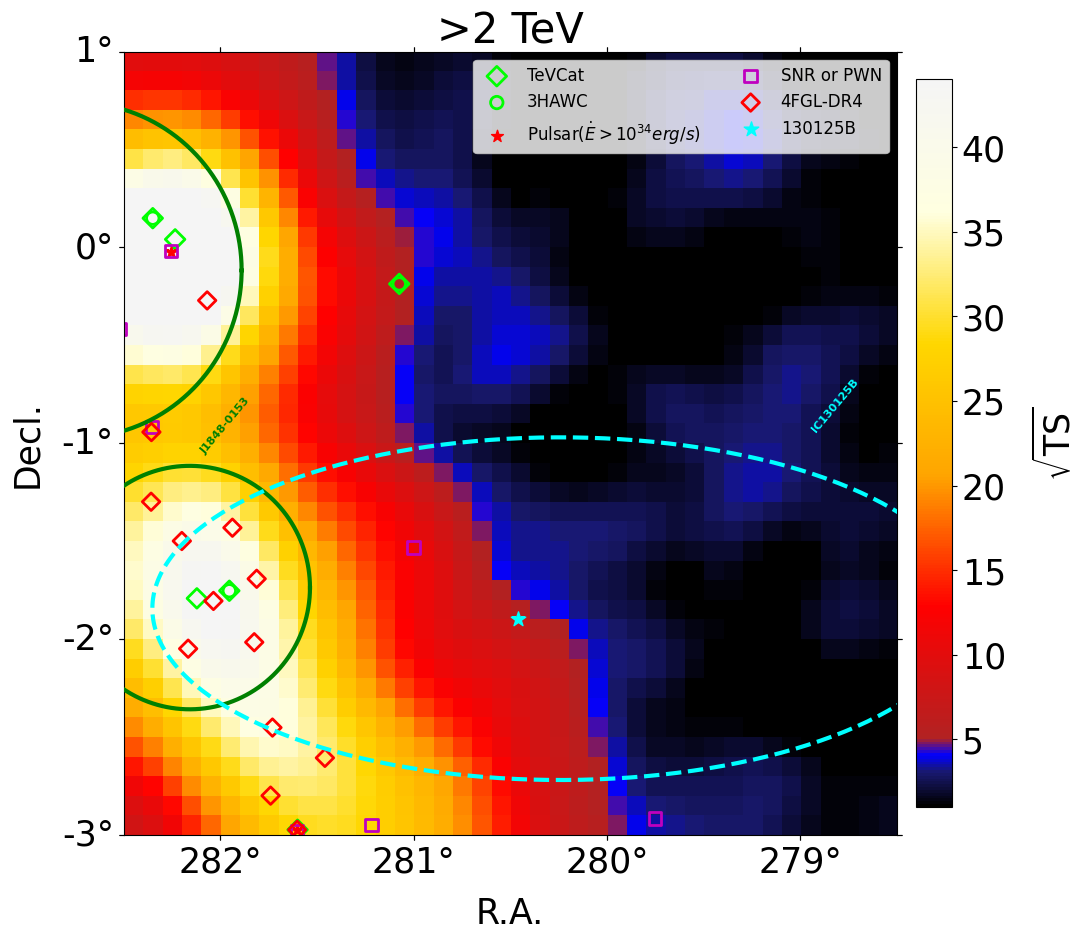}\\
\includegraphics[width=0.45\textwidth]{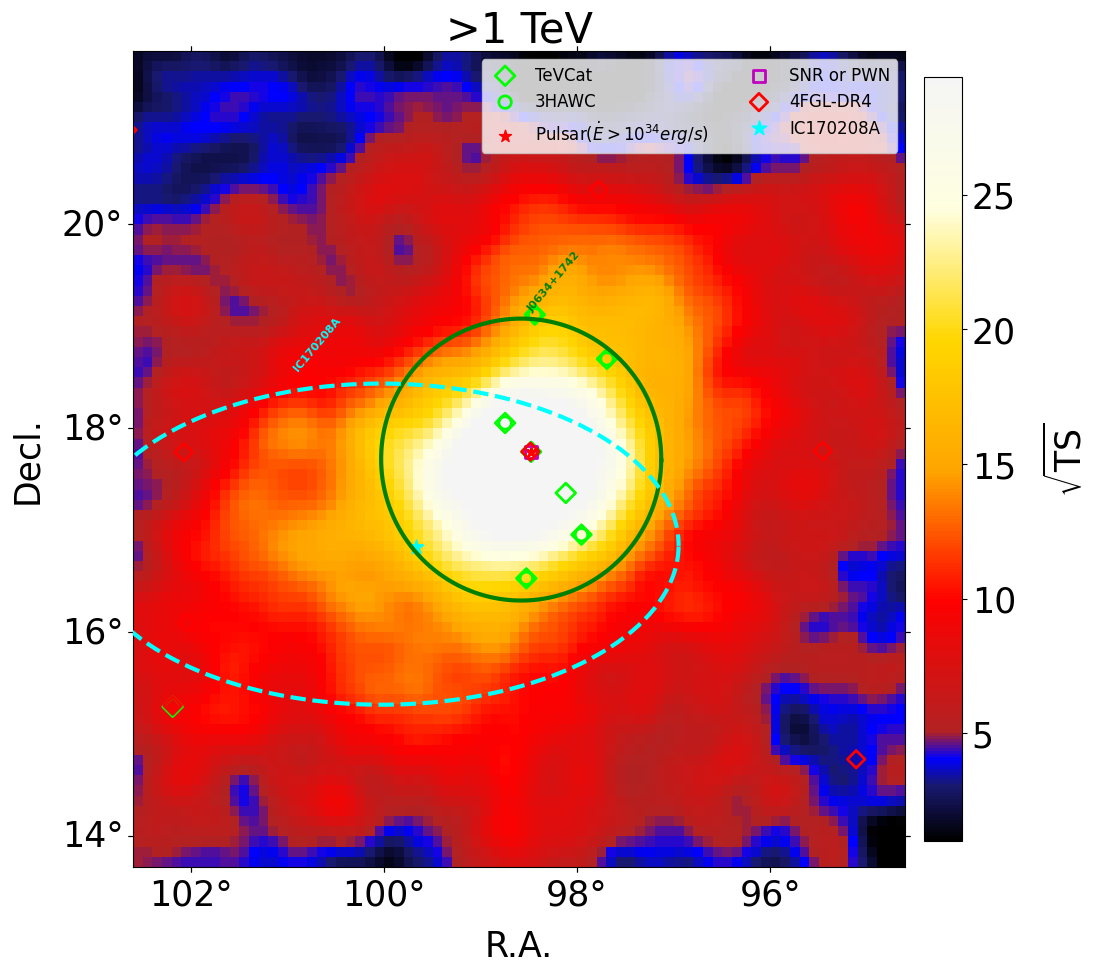}
\includegraphics[width=0.45\textwidth]{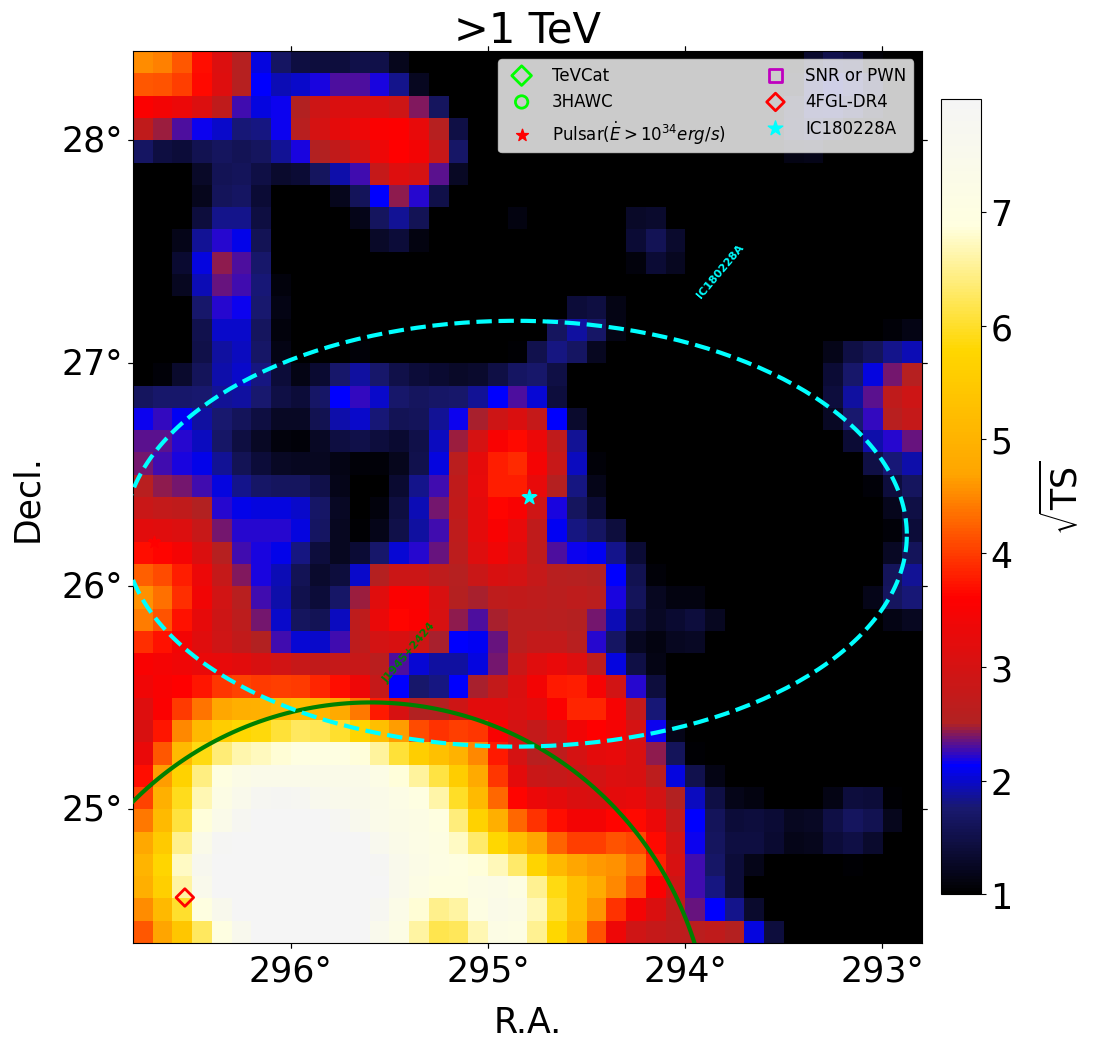}\\
\includegraphics[width=0.45\textwidth]{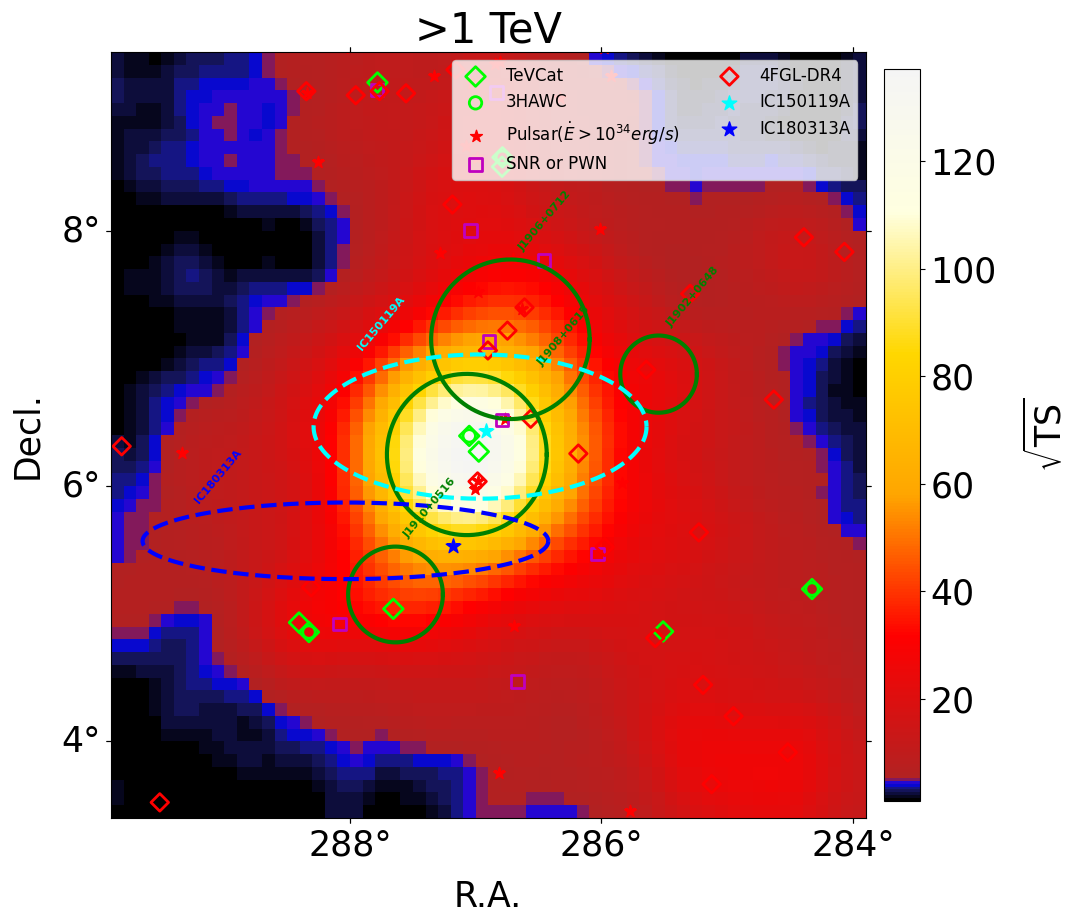}
\includegraphics[width=0.45\textwidth]{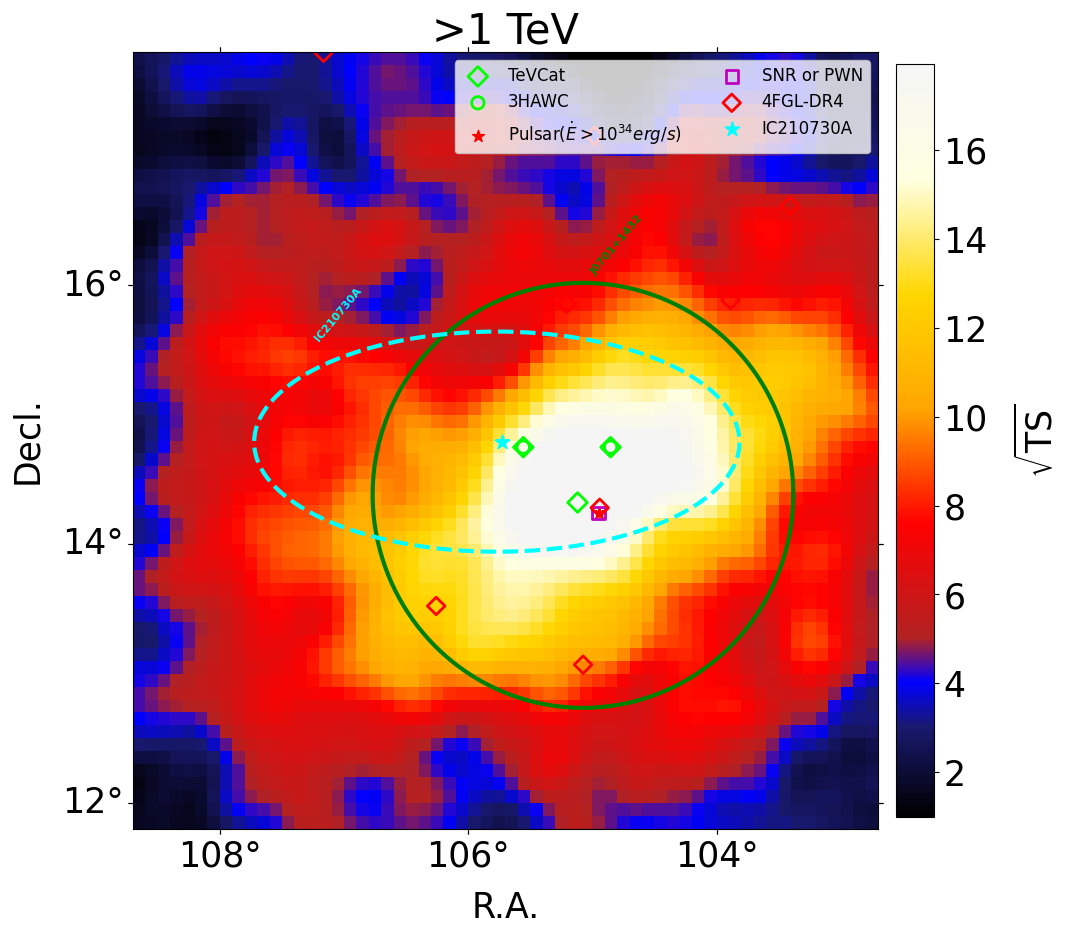}
\caption{The figure shows the significance maps of LHAASO sources that present the potential spatial coincidences with neutrino events. The solid green circles represent the approximate region of LHAASO sources by the joint fitting of WCDA and KM2A. Note that the green circle is a schematic illustration of the source and presents the approximate location, which may slightly deviate from the precise positions listed in Table~\ref{tab:tab1}. Two energy thresholds are applied: $>1\ \mathrm{TeV}$ for most sources and $>2\ \mathrm{TeV}$ for J1848-0153u due to air shower attenuation at large zenith angles. The dashed cyan ellipses mark the uncertainty region of alert events with the $90\%$ systematic uncertainty. Positions of sources in the catalog of other detections, such as HAWC \citep{abeysekara2017extended} and Fermi-LAT \citep{atwood2009large}, are marked on the sky map.}
\label{fig:skymap}
\end{figure*}

\begin{table*}[!htbp]
	\centering
	\caption{Spatial relationship of alert neutrino events (with $\Omega_{90}<20\,\rm deg^2$) and sources.}
	\begin{tabular}{cccccccccccc}
		\hline
		Event ID$^a$&RA$_{\rm alert}$& DEC$_{\rm alert}$ & $\Omega_{90}(\rm deg^2)$ & Source Name$^b$& RA$_{\rm source}$ & DEC$_{\rm source}$ & $\sigma_{p,95}^c$ & $r_{39}^d$ & Component & overlap ratio$^e$\\
		\hline
		IC111209A & $99.98^{+1.19}_{-2.02}$ & $20.42^{+1.6}_{-2.02}$ & 9.121857 & J0634+1741u & 98.57 & 17.69 & 0.10 & $0.89 \pm 0.04$ & KM2A/WCDA & 0.000074 \\
        \hline
        IC130125B & $280.46^{+1.89}_{-2.33}$ & $-1.9^{+0.93}_{-0.82}$ & 5.797225 & J1848-0153u & 282.06 & -1.89 & 0.10 & $0.51 \pm 0.03$ & WCDA/KM2A & 0.108300 \\
        \hline
        IC150119A & $286.92^{+1.36}_{-1.27}$ & $6.43^{+0.6}_{0.53}$ & 2.3329415 & J1902+0648 & 285.58 & 6.80 & 0.10 & $<0.15$ & WCDA & 0.009750\\
        \hline
        IC150119A & $286.92^{+1.36}_{-1.27}$ & $6.43^{+0.6}_{0.53}$ & 2.3329415 & J1906+0712 & 286.56 & 7.20 & 0.21 & $0.21\pm0.05$ & WCDA & 0.021530\\
        \hline
        IC150119A & $286.92^{+1.36}_{-1.27}$ & $6.43^{+0.6}_{0.53}$ & 2.3329415 & J1908+0615u & 287.05 & 6.26 & 0.05 & $0.43\pm0.02$ & KM2A/WCDA & 0.184736\\
        \hline
        IC170208A & $99.67^{+2.59}_{-3.3}$ & $16.84^{+1.6}_{-1.55}$ & 14.5644975 & J0634+1741u & 98.57 & 17.69 & 0.10 & $0.89 \pm 0.04$ & KM2A/WCDA & 0.172821\\
        \hline
        IC180228A & $294.79^{+1.85}_{-1.71}$ & $26.4^{+0.79}_{-1.12}$ & 5.337686 & J1945+2424 & 296.36 & 24.40 & 0.37 & $1.29 \pm 0.11$ & WCDA/KM2A & 0.050885\\
        \hline
        IC180313A & $287.18^{+0.75}_{-2.46}$ & $5.53^{+0.34}_{-0.26}$ & 1.51191 & J1908+0615u & 287.05 & 6.26 & 0.03 & $0.36\pm0.01$ & KM2A/WCDA & 0.001079\\
        \hline
        IC180313A & $287.18^{+0.75}_{-2.46}$ & $5.53^{+0.34}_{-0.26}$& 1.51191 & J1910+0516 & 287.55 & 5.28 & 0.15 & $<0.30$ & KM2A/WCDA & 0.072724\\
        \hline
        IC210730A & $105.73^{+1.93}_{-1.85}$ & $14.79^{+0.85}_{-0.85}$ & 5.04441 & J0703+1405 & 105.83 & 14.10 & 0.26 & $1.88\pm0.09$ & KM2A/WCDA & 0.989396\\
        \hline
	\end{tabular}
        \begin{tablenotes}
		  \item$^a$ IceCube alerts adopt the format IceCube-YYMMDDX. The string YYMMDD indicates the detection date, and the trailing uppercase letter indicates the sequential alert on a given day. 	
            \item$^b$ The source with a label “u” at the end of the name is the UHE source with TS$_{100} > 20$.
			\item$^c$ The $\sigma_{p,95}$ represents
            95\% statistical uncertainty of the source component position.
			\item$^d$ The $r_{39}$ is the 39\% containment radius of the 2D-Gaussian model. For
            the point-like source, the $95\%$ statistical upper limits are shown.
            \item$^e$ The overlap ratio represents the ratio of the overlapping area to the uncertain region of the alert neutrino event.
		\end{tablenotes}
	\label{tab:tab1}
\end{table*}

The results of this search are presented in Figure \ref{fig:skymap} and Table \ref{tab:tab1}. We use two energy thresholds for the plots: $>2\ \mathrm{TeV}$ for J1848-0153u and $>1\ \mathrm{TeV}$ for other sources. Due to the large zenith angle of J1848-0153u, the air shower induced by this source suffers more energy attenuation during atmospheric propagation, which increases the minimum reconstruction energy of WCDA. The significance TS maps in Figure \ref{fig:skymap} are produced from combined observational data of KM2A and WCDA. The green circles are fitted with joint data from WCDA and KM2A, which only serve as a schematic illustration of source morphology and approximate positions, rather than precise positions. For this reason, the source locations shown in the figure may differ slightly from the accurate positional parameters listed in Table \ref{tab:tab1}. In summary, nine spatial coincidences for the threshold of $\Omega_{90}\le 10\,\rm deg^2$, and ten for $\Omega_{90}\le 15\,\rm deg^2$ and $\Omega_{90}\le 20\,\rm deg^2$ between LHAASO sources and neutrino events are found, including seven neutrino events that are spatially coupled to eight LHAASO sources. Within the $90\%$ uncertainty regions of IC150119A and IC180313A, there are three and two sources, respectively. Figure~\ref{fig:skymap}, using WCDA data spanning 508 days and KM2A data spanning 933 days, displays significance maps for LHAASO sources that overlap with the $90\%$ uncertainty regions of neutrino events with energies $\geq$ 100 TeV, highlighting the spatial alignment between candidate sources and neutrino localization. Table~\ref{tab:tab1} summarizes the key properties of these overlapping pairs, including the neutrino event’s coordinates, the corresponding LHAASO source’s position and spatial parameters, and the detector(s) (KM2A/WCDA) that detected the source. The overlap ratio represents the ratio of the overlapping area to the uncertain region of the alert neutrino event. Note that we have retained those spatial 
coincidences with very small overlaps.  We applied the same rules in the Monte Carlo simulation process. Therefore, this treatment will not affect the stability of the results. Furthermore, since neutrino events could originate from any location within the uncertainty region, even coincidence with a very small overlap ratio would be meaningful.

\section{Coincidence analysis}\label{sec:analysis}

By searching the samples, we have checked the correlation, and the next step is to analyze the spatial coincidences. In Section \ref{sec:sample}, we mentioned that we have made some cuts in the IceCube catalog. According to $r_{90}$ of the IceCube alert events, we can calculate $\Omega_{90}$ for the alert events \citep{kouch2024association}.
Following the initial correlation search, we further analyzed the observed overlaps to distinguish true physical associations from random spatial coincidences. As noted in Section \ref{sec:sample}, our neutrino sample was filtered to include only events with $\Omega_{90} \leq 10~\rm deg^{2}$, $15~\rm deg^{2}$, $20~\rm deg^{2}$ and energy $\geq$100 TeV, minimizing the impact of poorly localized events on coincidence assessments. The $\Omega_{90}$ value for each neutrino event was calculated from its directional errors ($\rm RA^+, ~RA^-,~DEC^+,~DEC^-$) using the formula proposed by \citep{kouch2024association} shown as

\begin{equation}\label{equ:omega90}
\begin{aligned}
\Omega_{90} = \frac{\pi}{4}(\rm RA^+ \times DEC^+ + RA^+ \times DEC^- \\ \rm + RA^- \times DEC^+ + RA^- \times DEC^-).
\end{aligned}
\end{equation}
where the four parameters, $\rm RA^+$, $\rm RA^-$, $\rm DEC^+$, and $\rm DEC^-$, are directional errors \footnote{Here, we neglect the $ \cos(\rm DEC) $ projection for $ \rm RA^+$/$\rm RA^-$ errors, which can be important at high declination. However, LHAASO source sizes are computed below with the same method, so the spatial overlap statistic remains unbiased.}. In this work, a $90\%$ uncertainty radius is employed.
We obtained spatial coincidences through a Monte Carlo simulation. The simulation process is as follows.

\begin{enumerate}
\item The position of the source is established on the basis of the data in the LHAASO catalog with a higher significance level. And $\sqrt{\sigma_{p,95}^2 + r_{39}^2}$ serves as the extension radius.
\item Random RA of the IceCube events, ensuring that the declinations and associated error regions remain unchanged, due to the sensitivity of IceCube only depending on the zenith angle \citep{aartsen2017all}.
\item Repeat $N_{\rm MC}$ = 10,000 times and record the count of neutrinos that overlap with the source in each simulation. Denoted as $\nu_{\rm sim}$.
\item Count when the value is equal to or larger than the real one. Denoted as $N(\nu_{\rm sim}\geq\nu_{\rm real})$. Calculate the $p$-value as the probability of accidental coincidence \citep{davison1997bootstrap},
\end{enumerate}

\begin{equation}\label{equ:pvalue}
p = \frac{N(\nu_{\rm sim}\geq\nu_{\rm real})+1}{N_{\rm MC}+1}.
\end{equation}
The randomization of neutrino events is a process that is not influenced by the sources. Under this premise, we get the distribution of the test statistics of $\nu_{\rm sim}$ with the hypothesis that the source is not associated with neutrinos.

\begin{figure*}
\centering
\includegraphics[width=0.4\textwidth]{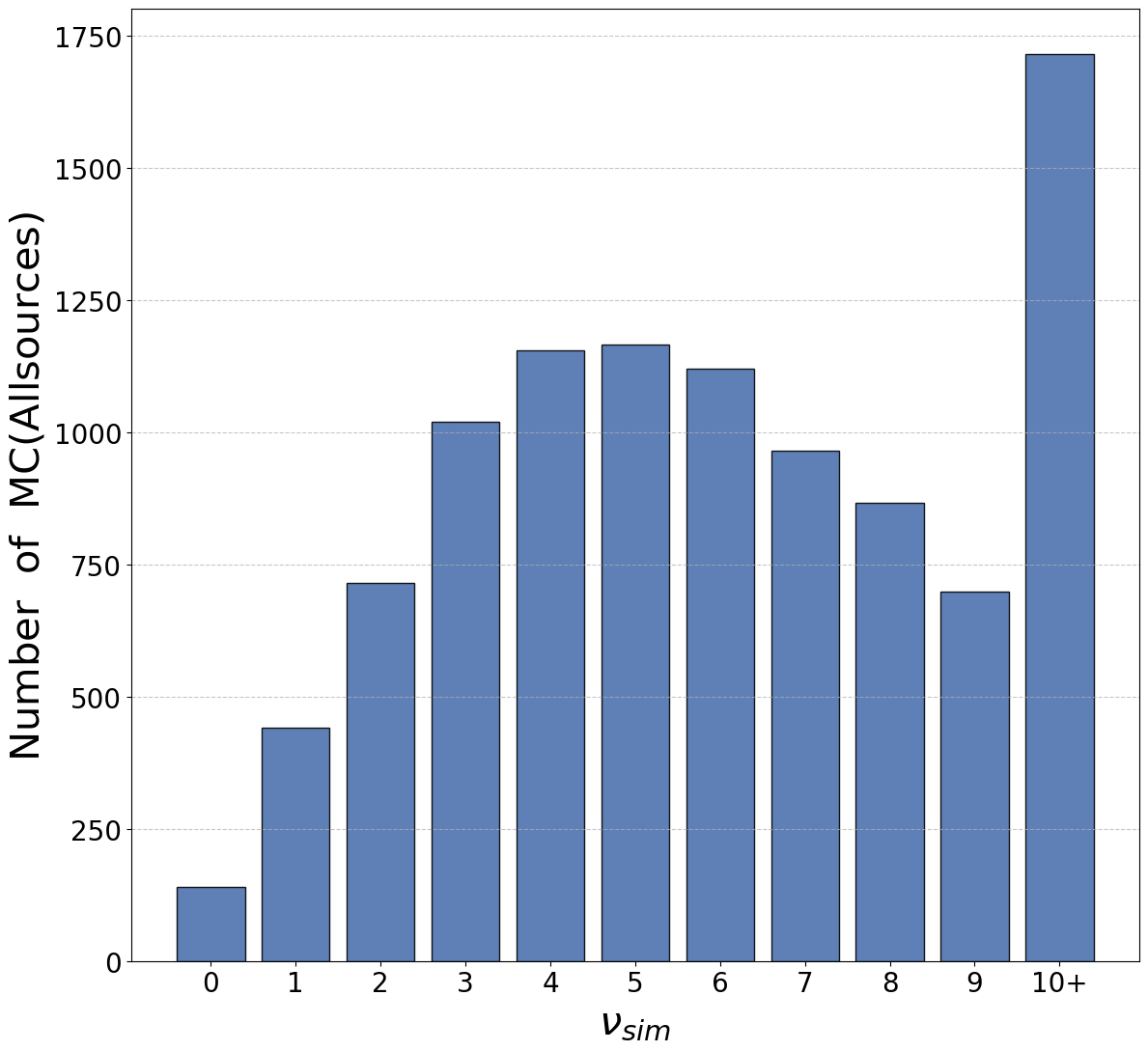}{(a)}
\includegraphics[width=0.4\textwidth]{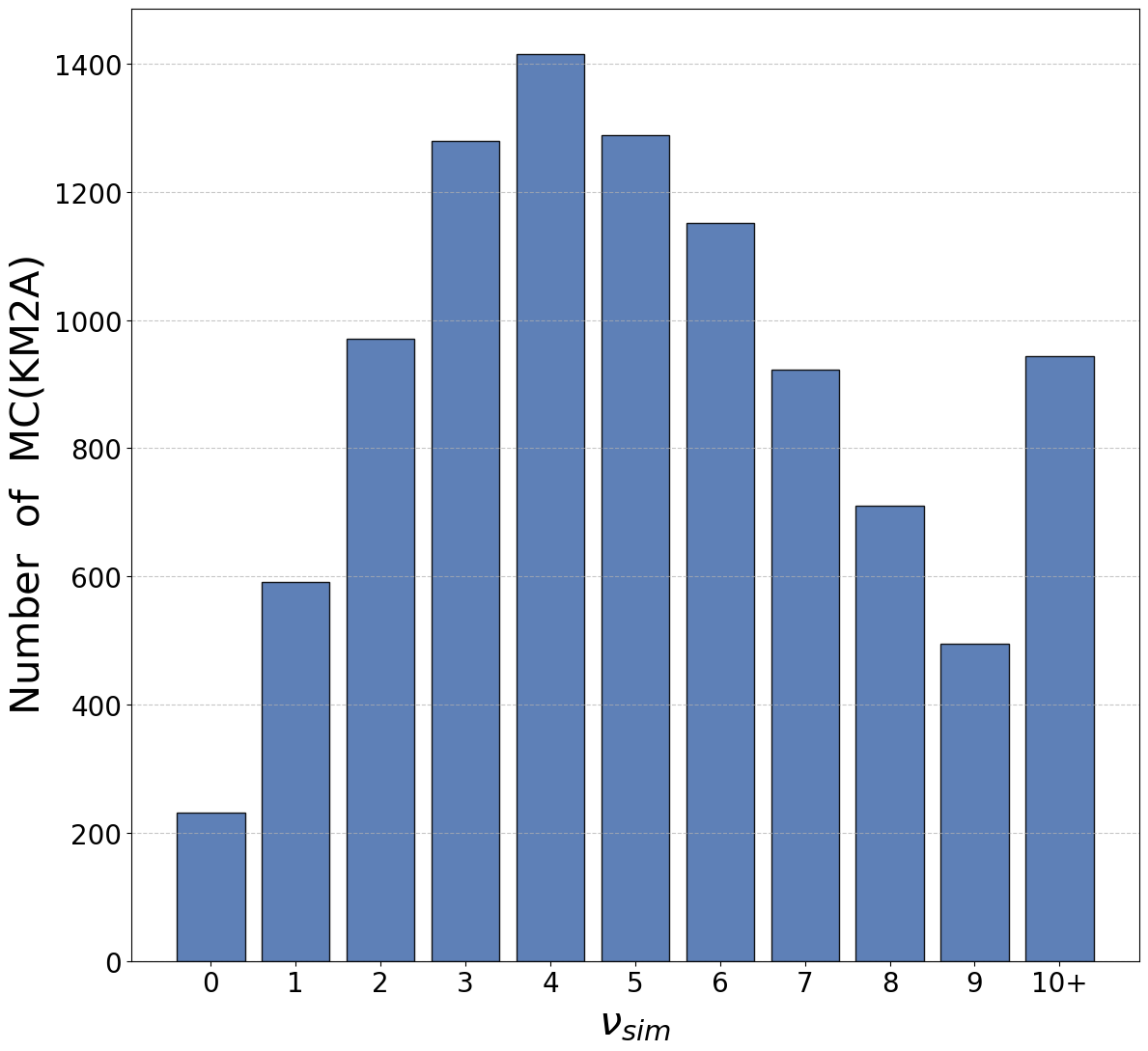}{(b)}
\includegraphics[width=0.4\textwidth]{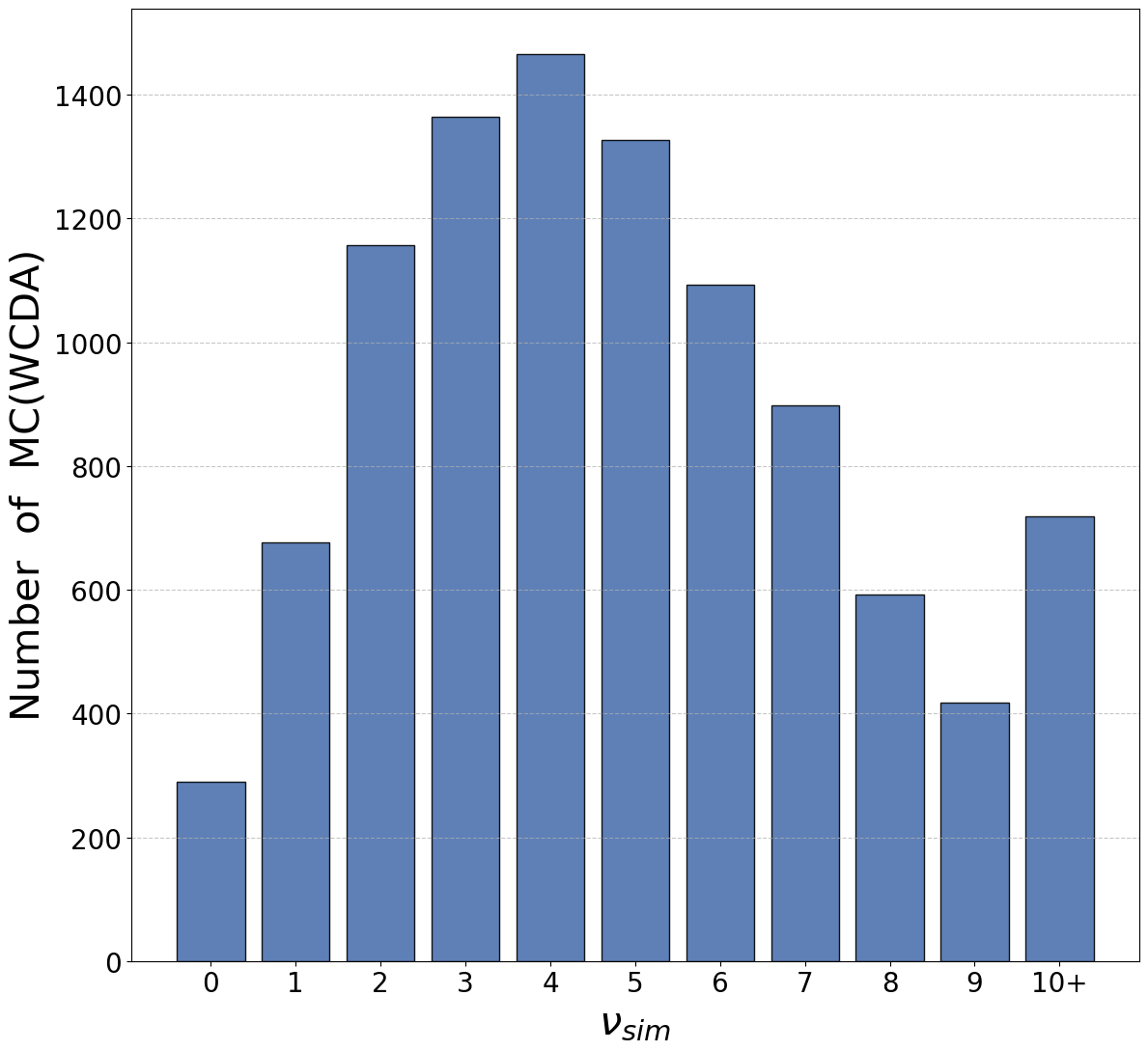}{(c)}
\caption{Results of 10,000 MC simulations with the cut ICECAT-1 sample ($\Omega_{90}\leq10\rm deg^2$ and Energy $\geq \rm 100TeV$). (a) All LHAASO sample: $\nu_{\rm real}=9$ (b) KM2A sample: $\nu_{\rm real}=7$ (c) WCDA sample: $\nu_{\rm real}=9$.}
\label{fig:MC}
\end{figure*}

We categorize LHAASO sources into three samples for statistical analysis: (I) the full sample including all detected sources; (II) sources only detected by KM2A; (III) sources only detected by WCDA. We adopt three thresholds for the $\Omega_{90}$: $\Omega_{90}\leq 10\rm \deg^2$, $\Omega_{90}\leq 15\rm \deg^2$ and $\Omega_{90}\leq 20\rm \deg^2$. The criterion $\Omega_{90}\leq 10\rm \deg^2$ is defined as our baseline selection, and the corresponding distributions from 10,000 Monte Carlo (MC) simulations are presented in Figure \ref{fig:MC}. The full results for all three thresholds are summarized in Table \ref{tab:MC}. For the baseline case $\Omega_{90}\leq 10\rm \deg^2$, the derived p-values are 0.237 for the full source sample, 0.312 for the KM2A-only sample, and 0.115 for the WCDA-only sample, with the observed coincidence counts $\nu_{\rm real}=9$, 7, and 9, respectively. For $\Omega_{90}\leq 15\rm \deg^2$ and $\Omega_{90}\leq 20\rm \deg^2$, the p-values range from 0.221 to 0.57. Across all three cuts, the overall trend remains consistent, which demonstrates that the p-value only slightly depends on the selection of $\Omega_{90}$. It should be noted that the three source samples are nested and overlapping, and their p-values correspond to non-independent statistical tests. All derived $p$-values are greater than 0.05. Statistically, we cannot reject the null hypothesis of random spatial coincidence, and no significant physical correlation is found. Previous studies have shown that changes in $\nu_{\rm real}$ or sample cut can affect numerical results \citep{plavin2020observational}. Especially when the source is a transient source, adding the time factor in the analysis can obtain better results. Unfortunately, most of the LHAASO sources are stable, and the time factor does not participate in our simulation. 

It is necessary to address the interference from Galactic diffuse emission for sources near the Galactic plane. The majority of LHAASO sources lie within $|b|<12^\circ$, a region where the well-established Galactic diffuse neutrino emission is prominent \citep{icecube2023observation}. The neutrino event displayed in the upper-right panel of Figure 1 is situated on the Galactic plane, so its signal may receive contributions from both discrete LHAASO sources and the large-scale diffuse neutrino component. According to the diffuse template analysis in Ref\citep{li2024search}, spatial overlaps between neutrino events and VHE gamma-ray sources in the Galactic plane cannot be directly interpreted as physical source associations. The mixing of discrete sources and diffuse emission weakens the reliability of these coincident pairs, which is another reason why we find no compelling evidence for a tight connection between LHAASO sources and IceCube neutrino alerts.

\begin{table*}[!htbp]
	\centering
	\caption{$p$-value of the source-neutrino correlation for different gamma-ray source samples.}

	\begin{tabular}{|c|c|c|}
		\hline
	    $\boldsymbol{\Omega_{90} \leq 10\,\mathrm{deg}^2}$& \text{$p$-value} & \text{$\nu_{\rm real}$} \\
		\hline
		 All sources & 0.237 & 9 \\
         KM2A sources & 0.312 & 7 \\
         WCDA sources & 0.115 & 9 \\
        \hline
	\end{tabular}
	
	\vspace{1em} 

	\begin{tabular}{|c|c|c|}
		\hline
		$\boldsymbol{\Omega_{90} \leq 15\,\mathrm{deg}^2}$& \text{$p$-value} & \text{$\nu_{\rm real}$} \\
		\hline
		 All sources & 0.41 & 10 \\
         KM2A sources & 0.462 & 8 \\
         WCDA sources & 0.221 & 10 \\
        \hline
	\end{tabular}
	
	\vspace{1em} 

	\begin{tabular}{|c|c|c|}
		\hline
		$\boldsymbol{\Omega_{90} \leq 20\,\mathrm{deg}^2}$& \text{$p$-value} & \text{$\nu_{\rm real}$} \\
		\hline
		 All sources & 0.534 & 10 \\
         KM2A sources & 0.57 & 8 \\
         WCDA sources & 0.326 & 10 \\
        \hline
	\end{tabular}
    
	\label{tab:MC}
\end{table*}

\section{Spectral analysis}

\begin{figure*}
\centering
\includegraphics[width=0.3\textwidth]{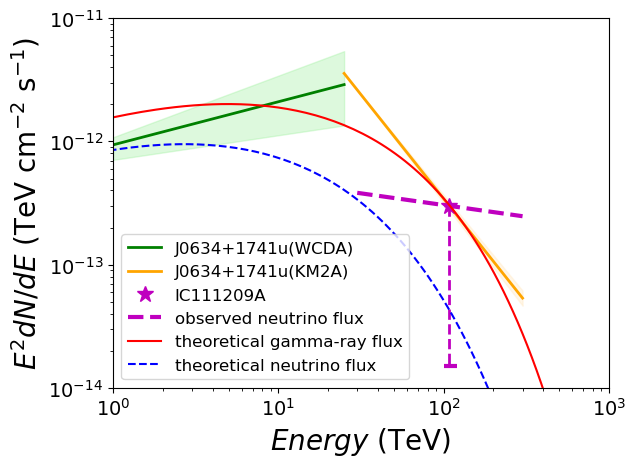}
\includegraphics[width=0.3\textwidth]{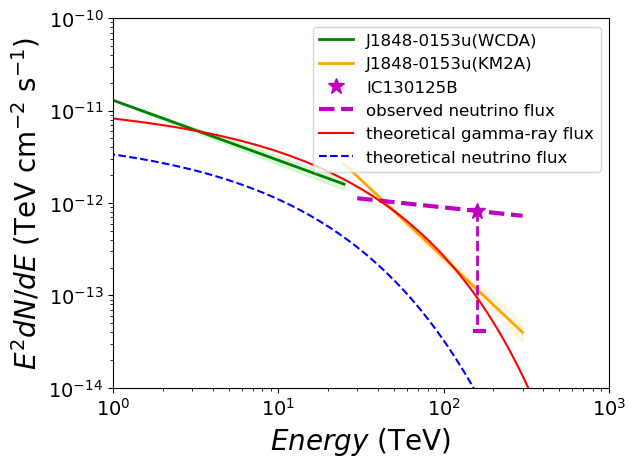}
\includegraphics[width=0.3\textwidth]{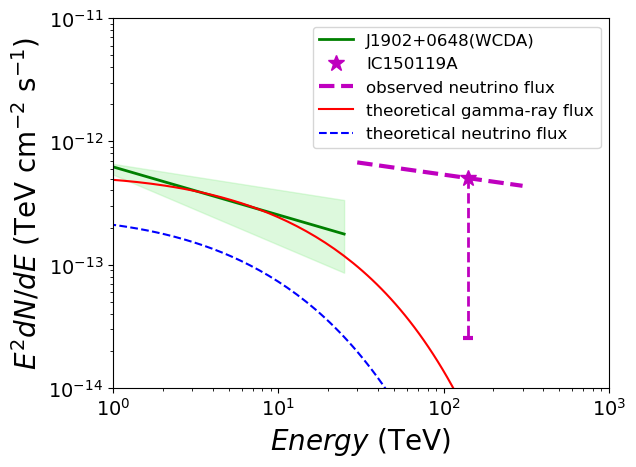}
\includegraphics[width=0.3\textwidth]{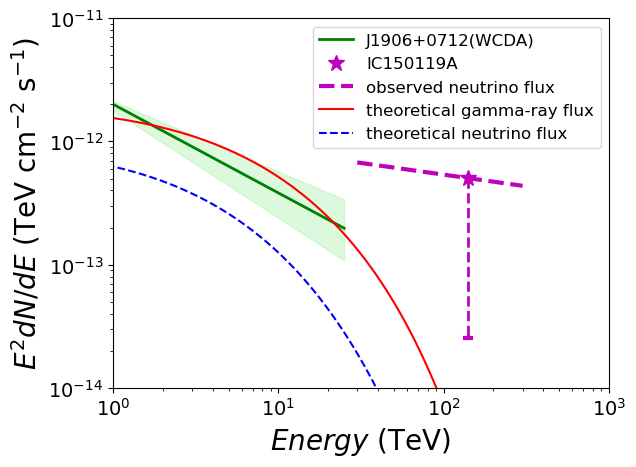}
\includegraphics[width=0.3\textwidth]{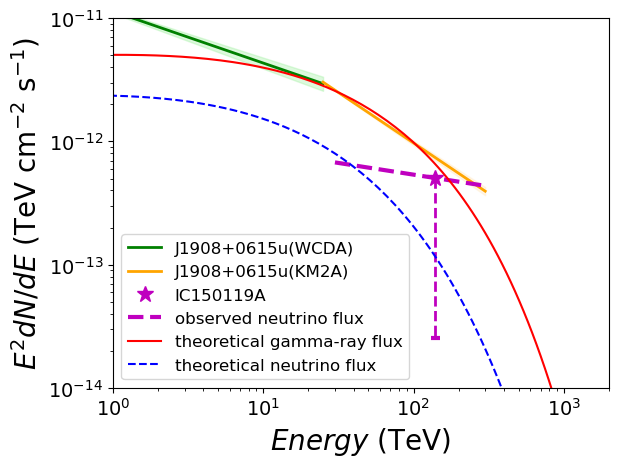}
\includegraphics[width=0.3\textwidth]{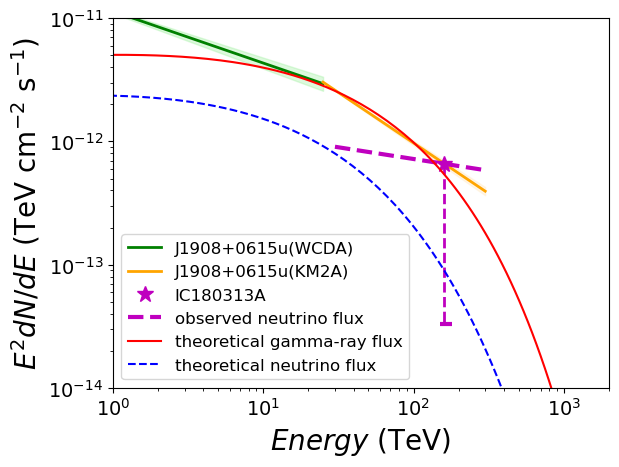}
\includegraphics[width=0.3\textwidth]{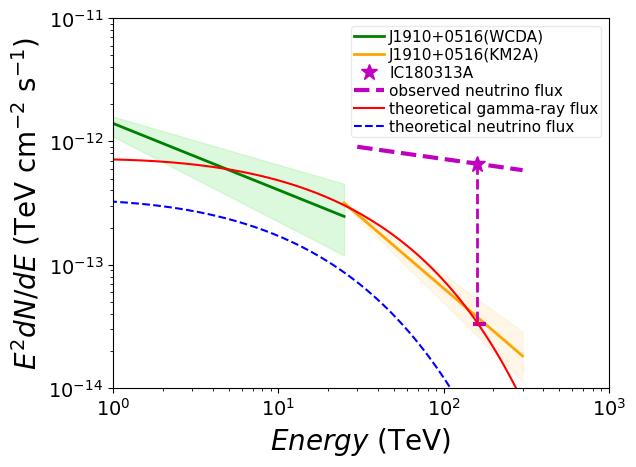}
\includegraphics[width=0.3\textwidth]{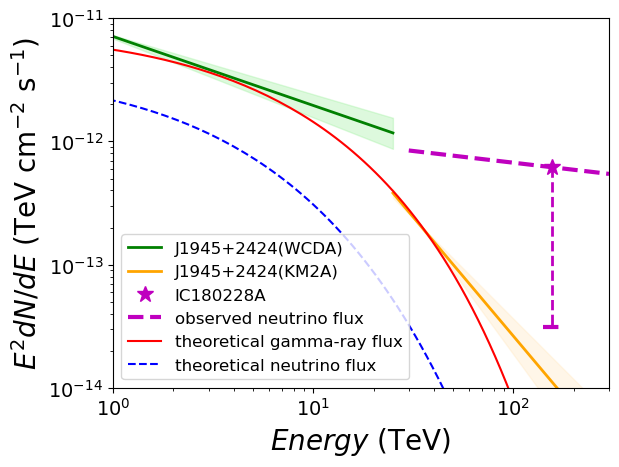}
\includegraphics[width=0.3\textwidth]{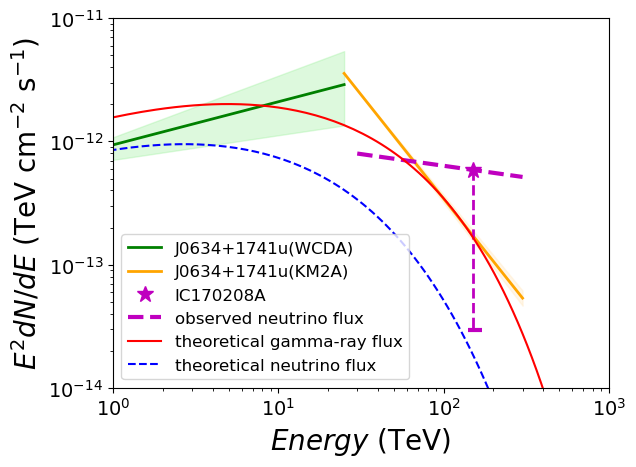}
\includegraphics[width=0.3\textwidth]{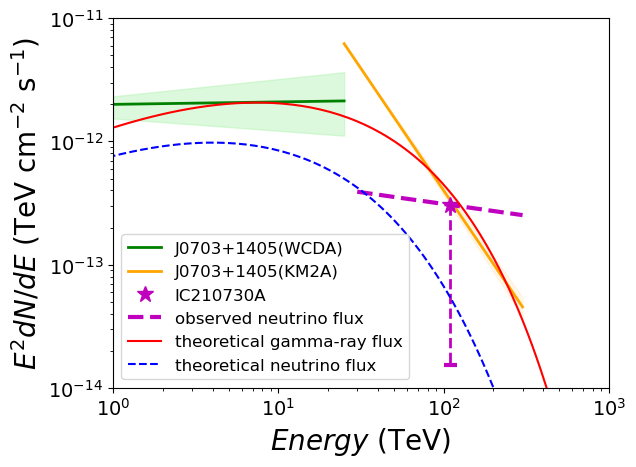}
\caption{The yellow solid line and the blue short-dashed line represent the flux of theoretical gamma rays and neutrinos from the $pp$ interaction, respectively. Straight lines and shaded areas represent the gamma-ray flux detected by LHAASO, including KM2A and WCDA~\citep{cao2024first}. The upper purple dashed line indicates the neutrino flux inferred from the observed event count in $[\rm 30\,TeV,300\,TeV]$, and the lower purple dashed line corresponds to the region of $95\%$ confidence level, and the purple star marks the true reconstructed energy of the corresponding neutrino event. The neutrino spectrum is calculated using the spectral index 2.19, consistent with the IceCat-1 standard. For the case of the purple star and dashed line above the theoretical neutrino flux (blue short-dashed line) from $pp$ interactions, it means this neutrino is not likely to be produced by this source.}
\label{fig:spectrum}
\end{figure*}

To further analyze the correlation between neutrinos and LHAASO sources, we analyzed the gamma-ray energy spectrum of these sources. For sources listed in Table~\ref{tab:tab1}, their gamma-ray energy fluxes detected by LHAASO are shown in Fig.~\ref{fig:spectrum}. The gamma-ray energy flux data were from ~\citep{cao2024first}. A power-law form is adopted to characterize the source spectrum, i.e., 
\begin{equation}\label{equ:PL}
\frac{dN}{dE}=N_{0}(\frac{E}{E_{0}})^{-\Gamma}.
\end{equation}
The normalization factor $N_0$ corresponds to the flux value at the reference energy $E_0$, which is defined differently for each dataset: 3 TeV for WCDA data and 50 TeV for KM2A data \citep{cao2024first}. The symbol $\Gamma$ in the equation refers to the spectral index.
We establish a spatial model of point sources and extended sources to fit the spatial position with high test significance, and take the detected current flux in this region as the gamma-ray energy flux. To assess whether the sources we researched could potentially be candidates for neutrino events, we assumed that the detected gamma rays are purely hadronic. Therefore, the most optimistic theoretical flux of neutrinos produced by the $pp$ interactions can be obtained (Typically, the subdominant $p\gamma$ interactions are neglected for the galactic sources). The distribution of protons is adopted as an exponential cut-off power-law, i.e.,
\begin{equation}\label{equ:proton}
\frac{dN_{p}}{dE_{p}}=\frac{A}{E^{\alpha}_{p}}{\rm exp}[-(\frac{E_{p}}{E_{\rm cut}})^{\beta}],
\end{equation}

\begin{equation}\label{equ:Etotal}
E_{\rm total} = \int E_p \,\frac{d N_p}{d E_p} dE_p,
\end{equation}
where A is the normalization factor, $E_p$ is the incident proton energy, and $E_{\rm cut}$ is the cutoff energy. The parameter values used during the spectral fittings for LHAASO sources are shown in Table~\ref{tab:parameter}. The spectral index $\alpha$ is derived by fitting LHAASO gamma-ray spectra, and $\beta$ is fixed to a typical value, i.e., $\beta=1.0$. A different value of $\beta$ will affect the profile of the exponential cutoff, whereas a typical $\beta=1.0$ is sufficient and fixed in the spectral fittings below. The total proton energy $E_{\rm total}$ and normalization factor A are calculated self-consistently to match the observed gamma-ray spectra by $pp$ interactions~\citep{kelner2006energy}, with the additional parameters including source distances and gas densities.

\begin{table*}
    \centering
    \caption{Parameters used during spectral fittings for LHAASO sources}
    \begin{tabular}{c c c c c c c c}
    \hline
    \hline
         Source name & Distance & $\theta$ &$\alpha$& $\beta$ & $E_{\rm total}$ & $E_{\rm cut}$&Asso\\
        & (kpc)&(deg)&       &  & (erg)&(TeV)&\\
        \hline
        J0634+1741u & $0.25^{(1)}$ & 1.19 & 1.5 & 1.0 & 5.18$\times10^{47}$  & 200 &Geminga\\
        J0703+1405 & $0.288^{(2)}$ & 1.88 & 1.22 & 1.0 & 5.8$\times10^{47}$  & 180 &2HWC J0700+143\\
        J1848-0153u & $5.3^{(3,4)}$ & 0.56 & 2.17 & 1.0 & 5.2$\times10^{49}$  & 200& HESS J1848-018\\ 
        J1902+0648 & 2.5 & 0.18 & 2.0 & 1.0 & 3.74$\times10^{49}$ & 140 & \\ 
        J1906+0712 & 2.5 & 0.29 & 2.0 & 1.0 & 1.32$\times10^{50}$ & 70 & \\ 
        J1908+0615u & $2.37^{(5)}$ & 0.43 & 2.0 & 1.0 & 3.23$\times10^{50}$ & 500 & MGRO J1908+06\\
        J1910+0516 & $5.5^{(6)}$ & 0.42 & 2.0 & 1.0 & 2.5$\times10^{50}$ & 300 & SS 433 w1\\
        J1945+2424 & 0.65 & 1.29 & 2.0 & 1.0 & 3.45$\times10^{49}$ & 50 & 2HWC J1949+244\\
        \hline
        
        \hline
    \end{tabular}
    \begin{tablenotes}
			\item The rightmost column lists the closest known TeV source counterpart within the region of the LHAASO source. The spectral parameter $\alpha$ is derived from fitting LHAASO gamma-ray spectra, and $\beta$ is fixed to 1.0 as a typical value. $E_{\mathrm{total}}$ denotes the total energy of accelerated parent protons, calculated from fitted spectra, source distances, and ambient gas densities. For J1848-0153, we set a gas density of 140 $\rm cm^{-3}$ \citep{cao2025observation}. For other sources, we set a typical
            interstellar medium density of 1 $\rm cm^{-3}$. (1)\cite{faherty2007trigonometric} (2)\citep{brisken2003distance} (3)\cite{chaves2008hess} (4)\cite{zhang2014parallax} (5)\cite{cao2024first} (6)\cite{abeysekara2018very}
			
    \end{tablenotes}
    \label{tab:parameter}
\end{table*}

For some of these sources, LHAASO J0634+1741u, LHAASO J1908+0615u, LHAASO J1848-0153u, LHAASO J1910+0516u, their distances from Earth have been determined by other detectors in the past literature~\citep{faherty2007trigonometric,chaves2008hess,abeysekara2018very,zhang2014parallax,cao2024first}. For LHAASO J1902+0648 and J1906+0712, since their position in the sky is in the adjacent area of LHAASO J1908+0615u ($\sim 2.37\,\rm kpc$~\citep{cao2024first}), we adopt a similar distance with a value of 2.5 kpc. For LHAASO J0703+1405, which is in the region of the Monogem pulsar halo, we adopt the same distance as Monogem~\citep{brisken2003distance}. For LHAASO J1945+2424, there is currently no specific observation of its distance; therefore, a distance of 0.65 kpc is adopted as an approximate value. In addition, the gas density of $140\,\rm cm^{-3}$ for J1848-0153 is adopted from the previous literature\citep{duvidovich2020radio,cao2025observation}, and for other sources, the gas density is unavailable, and a typical density of the interstellar medium is adopted, i.e., $1\,\rm cm^{-3}$. Consequently, the expected fluxes of gamma rays and neutrinos from $pp$ interactions can be calculated. Under the assumption that gamma rays are purely hadronic, the gamma-ray spectra detected by LHAASO are fitted in Fig.~\ref{fig:spectrum} and the adopted parameters are summarized in Table \ref{tab:parameter}. Note that several derived values \(E_{\mathrm{total}}\) slightly exceed the canonical cosmic-ray energy budget of a single supernova remnant $\sim10^{50}\ \mathrm{erg}$ (considering the acceleration efficiency $10\%$ for protons and the typical energy budget $10^{51}\ \mathrm{erg}$ for a supernova explosion). It may imply that the surrounding gas density is higher than the adopted typical ISM density or a larger proton acceleration efficiency. Specifically, a higher surrounding gas density will increase the $pp$ interaction efficiency and require a lower energy budget \(E_{\mathrm{total}}\). In practice, the specific parameters adopted in our calculation are not our primary concern. This is because, regardless of the parameter set employed, we arrive at nearly identical estimates for the most optimistic neutrino flux once the model is normalized to reproduce the observed gamma-ray emission. The core focus of the subsequent analysis, rather, is the comparison between this most optimistic neutrino flux and the IceCube data.

%our calculation is based on the pure $pp$ hadronic scenario, which represents an optimistic assumption. In reality, the gamma-ray emission from these sources may arise from a combination of hadronic and leptonic processes. As a result, the total proton energy estimated solely from pp interactions tends to overestimate the true energy budget of accelerated particles. Meanwhile, $E_{\mathrm{total}}$ scales inversely with ambient gas density, further demonstrating that the derived parameters are physically reasonable.

To evaluate the observational neutrino flux, we constrain the expected number of detected (anti-)muon neutrino events within the energy range($E\in[\rm 30\,TeV,300\,TeV]$) to the interval $[0.05, 1]$, This interval is derived based on Poisson statistics with a negligible atmospheric neutrino background for the high-energy alert sample:
The upper bound of 1 corresponds to one detected neutrino event in the selected energy window. The lower bound of 0.05 is the $95\%$ confidence level lower limit of the event count. Under the Poisson framework, this range yields a statistical significance of $\gtrsim 2\,\sigma$ for a single detected astrophysical neutrino event. The corresponding flux constraint is written as
\begin{equation}\label{equ:flux}
T_{\rm obs}\int_{\rm 30TeV}^{\rm 300TeV}\frac{dN_{\nu}}{dE_{\nu}} A_{\rm eff}(E_{\nu},\theta)dE_{\nu}=[0.05,1],
\end{equation}
where $E_{\nu}$ is the neutrino energy and $\theta$ is the zenith angle. The observation time ($T_{\rm obs}$) is the duration from the first running time of IceCube to October 1, 2023, the same as the duration of IceCat-1. The effective area ($A_{\rm eff}$) of IceCube, depending on the neutrino energy $E_{\nu}$, the zenith angle $\theta$, and the neutrino alert type, can be found in \citet{abbasi2023icecat}. In addition, a power-law neutrino spectrum is adopted in the calculations, namely, ${dN_{\nu}}/{dE_{\nu}}~\propto~E_{\nu}^{-2.19}$. We adopt a power-law spectral index of 2.19 to characterize the astrophysical neutrino spectrum. This value was derived by \citep{haack2018measurement} and serves as the baseline spectrum for the design of the IceCube real-time alert system updated in 2019, as well as for the calculation of signalness. We also performed tests with slightly varied spectral indices, and the main conclusions below remain unchanged. Therefore, the estimated average neutrino flux can be obtained from Eq.~\ref{equ:flux} and shown in Fig.~\ref{fig:spectrum}.

As Fig.~\ref{fig:spectrum} shows, the theoretical neutrino flux for most coincidences is lower than the observed neutrino flux, except for LHAASO J0634+1741u-IC111209A, J1908+0615u-IC150119A, and J1908+0615u-IC180313A. Our results still rely on the assumption that the gamma rays detected by LHAASO originate entirely from hadronic processes. If we instead allowed for a mixture of leptonic and hadronic contributions, the inferred neutrino flux would be lower, since leptonic processes do not produce neutrinos. This suggests that the observed coincidences are likely not robust, unless the neutrino detection is a statistical fluke.

\section{Summary}\label{sec:summary}

The sources detected by LHAASO are considered potential PeVatrons, as gamma rays with energy larger than 100 TeV have been observed from them. However, it is still uncertain whether these very high-energy (VHE) gamma rays come from leptonic or hadronic processes. High-energy neutrinos from galactic sources can be a smoking gun to judge the origin of galactic VHE gamma rays. In this work, we study the correlation between the LHAASO sources and the track-like neutrino alerts in IceCat-1. We find nine spatial coincidences for the threshold of $\Omega_{90}\le 10\,\rm deg^2$, and ten for $\Omega_{90}\le 15\,\rm deg^2$ and $\Omega_{90}\le 20\,\rm deg^2$, between LHAASO sources and high-energy neutrino events. Consider whether the spatial connection between the source and the event is just a coincidence. We employ MC simulation to calculate the possibility of spatial coincidence. By performing an MC simulation, we obtain a high $p$-value, suggesting that these coincidences may be accidental rather than physical currently. 

We summarize the main limitations of this analysis: (1) We adopt a simple overlap-counting statistic without utilizing angular resolution and spectral information, which makes our test relatively conservative; (2) Only IceCat-1 high-threshold alert events are used instead of the full IceCube track dataset, leading to reduced detection sensitivity; (3) Fixed energy cutoff and localization uncertainty $\Omega_{90}$ are applied before analysis; (4) The observational neutrino flux relies on a single power-law assumption with a index of $-2.19$, and the spectral index variations may affect the derived flux constraints. None of these limitations rule out the existence of genuine physical associations between LHAASO sources and high-energy neutrinos.

Specifically, in this work, we exclusively use the IceCat-1 alert track sample rather than the full IceCube track dataset. Our spatial correlation test adopts a simple overlap-counting test statistic, which does not incorporate per-event angular resolution and energy information used in unbinned likelihood point-source analyses. This approach is computationally straightforward but statistically conservative, which inevitably reduces the detection sensitivity. The IceCube alert sample has a higher energy threshold and fewer events, leading to obvious sensitivity loss compared to the full dataset. Therefore, the weak correlation we observed cannot simply be interpreted as a definitive absence of physical association between the LHAASO sources and high-energy neutrinos. It may also result from the limited statistics and high threshold of the IceCube alert-only sample. Future analyses using the complete IceCube event sample will help further verify the source-neutrino connection. To further explore the association between LHAASO sources and high-energy neutrinos, under the fully hadronic emission assumption of gamma rays, we compared the theoretical and most optimistic neutrino fluxes of LHAASO sources with the estimated neutrino fluxes based on IceCat-1 alert track sample. For most sources, the theoretical and most optimistic neutrino flux is lower than the observed flux based on the IceCat-1 alert track sample, indicating that the maximum neutrino flux from these sources is not enough to account for the detected events. If the full IceCube track dataset is used, the flux upper limit will be further tightened based on the more sensitive dataset.

%In addition, by assuming that the VHE gamma rays detected by LHAASO are purely hadronic, we compare the theoretical neutrino flux from the hadronic process with the actual neutrino flux detected by IceCube. The results of the spectrum fittings indicate that these associations are likely not robust. \textbf{In addition, we summarize the main limitations of this analysis: (1) We adopt a simple overlap-counting statistic without utilizing angular resolution and spectral information, which makes our test relatively conservative; (2) Only IceCat-1 high-threshold alert events are used instead of the full IceCube track dataset, leading to reduced detection sensitivity; (3) Fixed hard cuts on neutrino energy and localization uncertainty are applied before analysis; (4) The theoretical neutrino flux calculation relies on a single power-law spectral assumption, and slight spectral index variations may affect the derived flux constraints. None of these limitations rule out the existence of genuine physical associations between LHAASO sources and Galactic neutrinos.}

According to the current observational results, robust evidence of the connection between LHAASO sources and neutrino events is still lacking. In the previous study, the confidence level of IceCube and IceCube-Gen2 in detecting galactic neutrinos, based on the detection ability of the detector and the flux of the source, is estimated~\citep{2024ApJ...961..254S}, indicating that a larger neutrino detector is crucial for capturing high-energy neutrinos from galactic sources. The next-generation neutrino detectors, such as IceCube-Gen2~\citep{aartsen2021icecube}, HUNT~\citep{Huang:2023R8}, and TRIDENT~\citep{ye2023multi}, will significantly help us find the connection between galactic VHE sources and high-energy neutrinos in the future.

\begin{acknowledgments}
We thank Shao-qiang Xi, Zhuo Li, Tian-Qi Huang, Si-Ming Liu, Cong Li, Ruo-Yu Liu, Xiang-Yu Wang from the LHAASO Collaboration for their helpful comments, and the whole LHAASO Collaboration for the construction and operation of LHAASO. We also thank the anonymous reviewer for his/her constructive comments and suggestions. We acknowledge support from the National Natural Science Foundation of China under grant No.12003007 and the Fundamental Research Funds for the Central Universities (No. 2020kfyXJJS039).
\end{acknowledgments}

\section*{Data Availability} The data that support the findings of this article are openly available.
\section*{Interest Conflict} The authors declare that they have no conflict of interest.

\nocite{*}
\bibliographystyle{apsrev4-2}
\bibliography{reference}

\end{document}